\documentclass[english,apl, reprint, superscriptaddress]{revtex4-1}
\usepackage{mathptmx}
\usepackage[LGR,T1]{fontenc}
\usepackage[latin9]{inputenc}
\usepackage{amssymb}
\usepackage{graphicx}
\usepackage{subscript}

\makeatletter

\DeclareRobustCommand{\greektext}{%
  \fontencoding{LGR}\selectfont\def\encodingdefault{LGR}}
\DeclareRobustCommand{\textgreek}[1]{\leavevmode{\greektext #1}}
\ProvideTextCommand{\~}{LGR}[1]{\char126#1}

\makeatother

\usepackage{babel}
\begin{document}

\title{17,000 \%/W Second-Harmonic Conversion Efficiency in Single-Crystalline
Aluminum Nitride Microresonators}

\author{Alexander W. Bruch}

\affiliation{Department of Electrical Engineering, Yale University, New Haven,
CT 06511, USA}

\author{Xianwen Liu}

\affiliation{Department of Electrical Engineering, Yale University, New Haven,
CT 06511, USA}

\author{Xiang Guo}

\affiliation{Department of Electrical Engineering, Yale University, New Haven,
CT 06511, USA}

\author{Joshua B. Surya}

\affiliation{Department of Electrical Engineering, Yale University, New Haven,
CT 06511, USA}

\author{Zheng Gong}

\affiliation{Department of Electrical Engineering, Yale University, New Haven,
CT 06511, USA}

\author{Liang Zhang}

\affiliation{R\&D Center for Semiconductor Lighting, Institute of Semiconductors,
Chinese Academy of Sciences, Beijing 100083, China}

\author{Junxi Wang}

\affiliation{R\&D Center for Semiconductor Lighting, Institute of Semiconductors,
Chinese Academy of Sciences, Beijing 100083, China}

\author{Jianchang Yan}

\affiliation{R\&D Center for Semiconductor Lighting, Institute of Semiconductors,
Chinese Academy of Sciences, Beijing 100083, China}

\author{Hong X. Tang}
\email{hong.tang@yale.edu}

\affiliation{Department of Electrical Engineering, Yale University, New Haven,
CT 06511, USA}

\date{\today}
\begin{abstract}
High quality factor optical microcavities have been employed in a
variety of material systems to enhance nonlinear optical interactions.
While single-crystalline aluminum nitride microresonators have recently
emerged as a low loss platform for integrated nonlinear optics such
as four wave mixing and Raman lasing, few studies have investigated
this material for second-harmonic generation. In this Letter, we demonstrate
an optimized fabrication of dually-resonant phase-matched ring resonators
from epitaxial aluminum nitride thin films. An unprecendented second-harmonic
generation efficiency of 17,000\%/W is obtained in the low power regime
and pump depletion is observed at a relatively low input power of
3.5 mW. This poses epitaxial aluminum nitride as the highest efficiency
second-harmonic generator among current integrated platforms.
\end{abstract}
\maketitle
On-chip optical nonlinearity has gained a significant interest for
their applications in classical and quantum photonic information processing.
Compared to their bulk counterparts, the tight optical confinement
and facile device engineering in nanophotonic waveguides allow for
greater experimental control over common nonlinear processes such
as second-harmonic generation (SHG) \citep{Cazzanelli2011,Chang2016a,Furst2010,Gromovyi2017,Guo2016a,Lin2016,Wang2017,Wolf2017,Xiong2011,Roland2016a},
third harmonic generation (THG) \citep{Surya2018,Levy2011,Wang2016b},
and photon-pair generation \citep{Luo2017,Clemmen2009,Guo2016c,Sharping2006,Jin2014}.
These various nonlinear processes can be further combined, leading
to applications of broadening the span of frequency combs \citep{Guo2017a,Lin2016a,Xue2017,Kowligy2018},
f-2f referencing \citep{Hickstein2017}, and all-optical control of
energy transfer between optical modes \citep{Guo2018}. Optical ring
resonators have become a common modality to enhance the nonlinear
effects. To date, various integrated platforms such as lithium niobate
\citep{Luo2017,Wang2014,Wang2017,Jin2014,Chang2016a,Wolf2017,Lin2016a},
silicon nitride (SiN) \citep{Ning2012a,Levy2011,Ning2012,Wang2016b,Xue2017},
gallium nitride (GaN) \citep{Xiong2011,Gromovyi2017,Roland2016a},
and aluminum nitride (AlN) \citep{Guo2016a,Guo2017a,Pernice2012a}
have been employed to enhance the SHG efficiency beyond those seen
in bulk crystals.  AlN is particularly attractive for SHG for its
large material band gap of 6.2 eV and mature fabrication processing.
Polycrystalline AlN microring resonators \citep{Guo2016a} have previously
demonstrated state-of-the-art on-chip SHG conversion efficiency of
2500 \%/W (0.025 mW\textsuperscript{-1}), owing to careful engineering
of the phase-matching between the infrared and visible modes as well
as efficient waveguide coupling for both pump and SHG light. However,
the efficiency of this device was ultimately limited by the intrinsic
optical quality (\emph{Q}) factor at both infrared (IR) and visible
wavelengths. Increasing the material quality will ultimately reduce
the optical loss at both fundamental and SHG wavelengths, allowing
new record SHG efficiency.

Single-crystalline AlN has emerged as the likely successor to polycrystalline
AlN photonics. As noted in previous studies, AlN epitaxially grown
on sapphire substrates exhibit exceptionally high film quality and
can achieve optical \emph{Q} factors on the order of $2\mathrm{x}10^{6}$
in the telecom band \citep{Liu2017c}, much higher than those typically
observed in polycrystalline AlN \citep{Guo2016a,Surya2018}. Currently,
AlN-on-sapphire devices have been employed for linear and nonlinear
phenomena at telecom IR wavelengths \citep{Liu2017a,Liu2018,Liu2016d}
as well as in the ultraviolet and visible regimes \citep{Lu2018,Liu2017b}.
Recently demonstrated UV SHG \citep{Fanto2016,Troha2016} in AlN-on-sapphire
devices was only observable using a pulsed pump while visible SHG
in this platform \citep{Liu2017b} realized a low conversion efficiency
due to the phase-mismatched condition. Lithography and etching challenges
in processing thick (> 1 \textgreek{m}m) AlN-on-sapphire films have
proven as a barrier to achieving a fully-etched AlN waveguide \citep{Liu2016d,Liu2017a,Liu2017c,Liu2018},
which is necessary to confine the phase-matched higher-order visible
modes within the ring resonators. 

In this Letter, we demonstrate a fabrication process for realizing
fully-etched AlN-on-sapphire devices with a 1.1 \textgreek{m}m thickness.
This process is then employed to fabricate ring resonators with both
high optical \emph{Q }factors as well as a tunable phase-matching
bandwidth of 1480-1640 nm. By optimizing the visible extraction waveguide
and the coupling conditions of both infrared and visible modes, an
on-chip SHG conversion efficiency of 17000 \%/W was achieved in the
low-power regime and a $\chi^{(2)}$ value of $6.2\pm0.4$ pm/V is
extracted for epitaxial AlN. Pump depletion was observed in the high-power
regime, with an absolute conversion efficiency as high as 11\%.

We model the SHG process in a microring cavity possessing both an
infrared and visible mode at angular frequencies $\omega_{IR}$ and
$\omega_{vis}$, respectively, each with internal loss rate, $\kappa_{0,IR(vis)}$
and external coupling rate $\kappa_{c,IR(vis)}$. The coupling strength
between these two resonance modes is represented by the parameter
$g$, which can be determined from the mode overlap between infrared
and visible modes, the microring geometry, and the $\chi^{(2)}$ coefficient
of AlN \citep{Guo2016a}. When pumping at a certain angular frequency,
$\omega_{p}$, SHG efficiency in the non-depletion regime can be expressed
in terms of these cavity coupling rates as well as the detuning between
the infrared (visible) resonances from the pump, $\delta_{IR}=\omega_{IR}-\omega_{p}$
$(\delta_{vis}=\omega_{vis}-2\omega_{p})$, as

\begin{equation}
\eta_{SHG}=\frac{P_{SHG}}{P_{p}^{2}}=g^{2}\frac{\hbar\omega_{vis}}{(\hbar\omega_{p})^{2}}\frac{2\kappa_{c,vis}}{\delta_{vis}^{2}+\kappa_{vis}^{2}}(\frac{2\kappa_{c,IR}}{\delta_{IR}^{2}+\kappa_{IR}^{2}})^{2}
\end{equation}
where $P_{SHG}$ and $P_{p}$ represent the on-chip SHG and pump power,
respectively, and $\kappa_{IR(vis)}$ is the total coupling rate $\kappa_{IR(vis)}=\kappa_{c,IR(vis)}+\kappa_{0,IR(vis)}$.
Under ideal phase matching conditions, the microring geometry is optimized
such that $\delta_{IR}=\delta_{vis}=0$ and $\omega_{vis}=2\omega_{IR}=2\omega_{p}$.
Fulfilling this dual-resonance condition, the SHG conversion efficiency
can be reduced to

\begin{equation}
\eta_{SHG}=\frac{P_{SHG}}{P_{p}^{2}}=\frac{16g^{2}}{\hbar\omega_{p}}\frac{\kappa_{c,vis}}{(\kappa_{c,vis}+\kappa_{0,vis})^{2}}\frac{\kappa_{c,IR}^{2}}{(\kappa_{c,IR}+\kappa_{0,IR})^{4}}\label{eq:eta_full}
\end{equation}

According to Eq. \ref{eq:eta_full}, maximum $\eta_{SHG}$ can be
achieved when both the IR and visible modes are critically coupled
($\kappa_{c}=\kappa_{0}$) and the SHG efficiency is dependent only
on the intrinsic device parameters. As noted in our previous report
in polycrystalline AlN \citep{Guo2016a}, it is practically difficult
to simultaneously achieve critical coupling for both modes in the
phase-matched geometry and the maximum conversion efficiency is often
limited by the cavity coupling rate. 

In this experiment, single-crystalline AlN films are grown on c-plane
sapphire by metal-organic vapor deposition (MOCVD) to a nominal thickness
of 1.1 \textgreek{m}m. The AlN-on-sapphire platform is quite insulating
compared to typical silicon-based substrates and suffers from significant
charging effects in our 100 kV electron-beam lithography process (Raith
EBPG 5000+). While metal masks are typically used for increasing the
conductivity of insulating substrates, many are not compatible with
typical negative-tone resists such as hydrogen silsesquioxane (HSQ)
and/or require a two-step etching process that ultimately limits the
process selectivity \citep{Liu2017c}. To mitigate charging effects,
we first spin a conventional FOx-16 HSQ mask followed by 300 nm poly(4-styrenesulfonic
acid) (PSSA) before sputtering 10 nm gold. PSSA acts as a water-soluble
sacrificial layer between the conductive gold layer and FOx-16 resist,
allowing sufficient charge transfer to the HSQ layer and is easily
removed during the development process \citep{Rooks}. Patterns are
then developed in a 25\% TMAH solution to maintain a high contrast.
The patterns are then transferred to the AlN via an optimized Cl\textsubscript{2}/BCl\textsubscript{3}/Ar
inductively coupled plasma (ICP) etch with a selectivity of 2.7:1
between AlN and FOx-16. Devices are then encapsulated in 1.0 \textgreek{m}m
plasma-enhanced chemical vapor deposition (PECVD) SiO\textsubscript{2}
and cleaved prior to measurement.

This process provides a simplified route for AlN-on-sapphire fabrication,
requiring a single masking step compared to the two-step processes
previously reported in Ref. \citep{Liu2017c}. The thicker FOx-16
mask and the high etching selectivity allow up to 1.4 \textgreek{m}m
of AlN to be fully etched, compared to 1.0 \textgreek{m}m demonstrated
previously. This fully etched structure is particularly important
for tight confinement of the visible mode. Finite element method simulations
indicate that the second-harmonic mode exhibits 1.2 dB/cm radiation
loss when leaving a 200 nm unetched AlN layer (radius = 30um, height
= 1 \textgreek{m}m, width = 1.2 \textgreek{m}m), whereas the radiation
loss is negligible in the fully-etched structure.

\begin{figure}
\begin{centering}
\includegraphics[width=1\columnwidth]{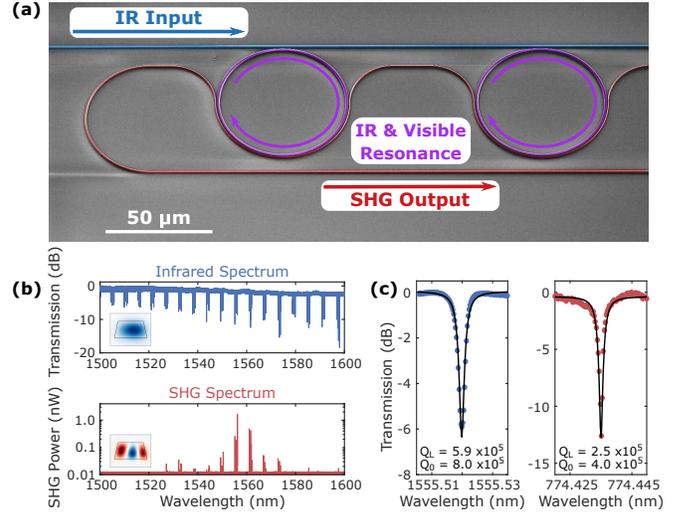}
\par\end{centering}
\caption{\label{fig:AlN-device-design}AlN device design and performance. (a)
Scanning electron microscope (SEM) micrograph of a fabricated AlN-on-sapphire
device before encapsulation in PECVD SiO\protect\textsubscript{2}.
A straight bus waveguide couples telecom IR light (blue) into cascaded
microring resonators exciting a dual resonance at infrared and visible
wavelengths (purple). The wrap around waveguide efficiently extracts
near-visible SHG light (red) from the microrings and passes the output
to an on-chip WDM \citep{Guo2016a,Guo2016b}. (b) Typical spectra
for phase-matched microring resonators. Sweeping an IR laser (blue)
shows multiple resonance dips corresponding to the cascaded microring
structure. Phase matched resonances near 1555-1560 nm can be identified
by peaks in SHG spectrum (red). The insets highlight the IR $\mathrm{TM_{0,0}}$
pump mode and the phase matched $\mathrm{TM_{2,0}}$ SHG mode, respectively.
(c) Resonance spectra of infrared $\mathrm{TM_{0,0}}$ and near-visible
$\mathrm{TM_{2,0}}$ modes with loaded and intrinsic \emph{Q} values
shown in the insets. }

\end{figure}

Fig. \ref{fig:AlN-device-design}(a) depicts the final fabricated
devices, where a 900 nm-wide bus waveguide is coupled to a microring
resonator with a waveguide width of 1.2-1.3 \textgreek{m}m. A thinner
waveguide wrapped around the ring is employed to efficiently extract
the near-visible light from the phase-matched ring resonator. The
near-visible extraction waveguide width is tapered from 175 to 125
nm at a constant coupling gap. As a result, the fundamental mode of
the extraction waveguide can be sufficiently matched to the near-visible
$\mathrm{TM_{2,0}}$ ring mode while minimally perturbing the infrared
$\mathrm{TM_{0,0}}$ ring mode.  Seven ring resonators share the
same IR bus and visible drop waveguides to account for fabrication
imperfections across the AlN chip. The infrared and visible waveguides
rejoin in an on-chip wavelength division multiplexer (WDM), while
an off-chip WDM separates the infrared and near-visible light for
simultaneous characterization \citep{Guo2016a,Guo2016b,Guo2017a}. 

Microring resonances are observed by sweeping the wavelength of an
infrared external cavity tunable diode laser (Santec TSL 710), as
shown in Fig. \ref{fig:AlN-device-design}(b). In phase-matched devices,
cavity resonances in the infrared spectrum occur simultaneously with
peaks observed in the corresponding visible spectrum, indicative of
the dual-resonance process. Fig. \ref{fig:AlN-device-design}(c) shows
typical resonance spectra at 1550 nm and 775 nm. The infrared resonance
is slightly undercoupled with a loaded \emph{Q }factor ($Q_{L}$)
of $5.9\mathrm{x}10^{5}$, while the second-harmonic mode is nearly
critically coupled with a $Q_{L}$ of $2.5\mathrm{x}10^{5}$. We note
that the extinction ratio of the infrared resonance increases when
scanning from shorter to longer wavelengths in Fig. \ref{fig:AlN-device-design}(b)
with minimal impact on the optical \emph{Q }factors. These optical
\emph{Q} factors are more than doubled compared to polycrystalline
AlN devices with a similar geometry \citep{Guo2016a} and are of similar
order of magnitude compared to previous AlN-on-sapphire devices \citep{Liu2017c},
which had larger dimensions and were partially etched, inducing lower
sidewall scattering loss but higher visible radiation loss as described
earlier.

\begin{figure}
\begin{centering}
\includegraphics[width=1\columnwidth]{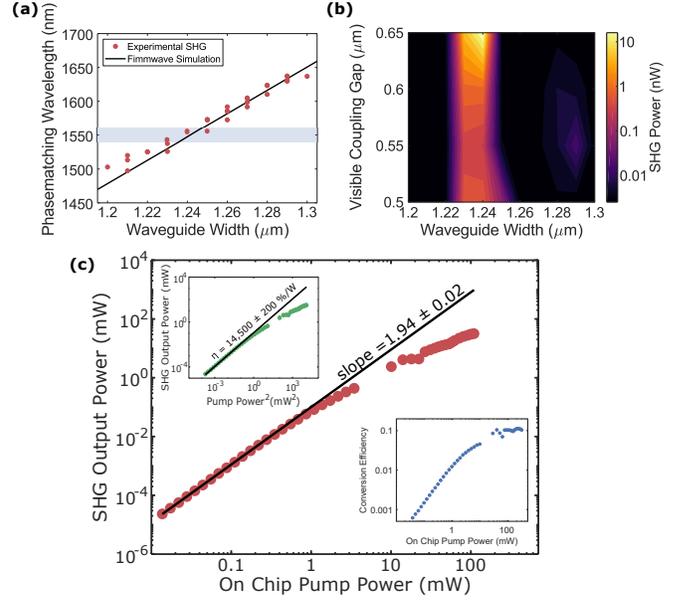}
\par\end{centering}
\caption{\label{fig:Geometrical-optimization}Geometrical optimization of SHG
performance. (a) Experimentally recorded and simulated SHG wavelength
for devices with ring widths of 1.2-1.3 \textgreek{m}m. The two data
sets exhibit a high degree of agreement. (b) Heat map of experimental
SHG power as a function of the waveguide width and the visible coupling
gap for the EDFA window (gray band in (a)). Maximum SHG power of 20
nW is observed for waveguide width 1.24 \textgreek{m}m and visible
coupling gap of 0.65 \textgreek{m}m. (c) Pump power dependence of
SHG output power. Experimental data (red) is plotted with a linear
fit of the low power region (solid line, slope = 1.94 $\pm$ 0.02).
The lower inset (blue) shows the absolute conversion efficiency over
the same pump power. The upper inset (green) shows the variation of
SHG output power versus $P_{p}^{2}$. A linear fit of the low power
region shows a SH conversion efficiency of $14,500\pm200$ \%/W.}
\end{figure}

We then optimize the SHG wavelength by varying the width of the AlN
ring waveguide, as shown in Fig. \ref{fig:Geometrical-optimization}(a).
By scanning the ring width from 1.2 to 1.3 \textgreek{m}m, SHG can
be observed over the entire range of our tunable laser (1480-1640
nm). The experimental SHG wavelength agrees remarkably well with the
calculated results from a FIMMWAVE simulation. We then optimize the
extraction efficiency at a particular wavelength by varying the coupling
gap of the visible extraction waveguide. Fig. \ref{fig:Geometrical-optimization}(b)
displays the SHG power collected as a function of both the phase-matching
width and visible coupling gap in the C-band erbium-doped fiber amplifier
(EDFA) window of 1540-1560 nm (gray band in Fig. \ref{fig:Geometrical-optimization}(a)).
The observed SHG power is very sensitive to the ring waveguide width,
which is the main determinant of the phase-matching condition between
the IR and visible modes. In contrast, a variety of visible coupling
gaps are able to extract the SHG power, due to the adoption of the
wrapped extraction waveguides. From an input pump power of 190 \textgreek{m}W,
a maximum external SHG power of 20 nW can be collected from a ring
waveguide width of 1.24 \textgreek{m}m and a visible coupling gap
of 0.65 \textgreek{m}m, whereas the SHG output decreases for visible
coupling gaps below 0.60 \textgreek{m}m. Scanning reference devices
at the SHG wavelength reveals that the 0.60-0.65 \textgreek{m}m gaps
are nearly critically coupled, whereas devices with visible coupling
gaps of 0.50-0.55 \textgreek{m}m are overcoupled and exhibit diminished
SHG output as predicted by Eq. \ref{eq:eta_full}.

At low pump levels, SHG output power is known to increase quadratically
with the pump power and deviates from this behavior with high pump
power due to pump depletion. Fig. \ref{fig:Geometrical-optimization}(c)
depicts the pump power dependence of the SHG output power in a device
with a 1.24 \textgreek{m}m width and a 0.65 \textgreek{m}m visible
coupling gap. At low pump power, the variation of the SHG output power
is close to the predicted quadratic dependence on the pump power,
with a fitted slope of 1.94 $\pm$0.02, whereas the SHG output power
begins to deviate from the quadratic behavior at high pump power.
The lower inset to Fig. \ref{fig:Geometrical-optimization}(c) shows
the absolute conversion efficiency ($P_{SHG}/P_{P}$), which deviates
greater than 50\% from the predicted behavior when the on-chip pump
power is greater than 3.5 mW and saturates to 11\% when $P_{P}$ is
greater than 35 mW. This deviation from the quadratic power dependence
is observed in the high power regime, which indicates significant
pump depletion and ultimately limits the maximum SHG efficiency \citep{Furst2010,Guo2016a}.
Looking to Eq. 1, we notice that $P_{SHG}$ varies linearly with $P_{p}^{2}$
with a slope equal to the SHG efficiency $\eta$; the upper inset
to Fig. \ref{fig:Geometrical-optimization}(c) shows this relation
and a SHG efficiency of $14,500\pm200$ \%/W for this device is found
via a linear fit of the low power regime. We then use the experimentally
measured values of $\kappa_{0}$ and $\kappa_{c}$ at infrared and
near-visible regimes to derive a nonlinear coupling strength $g/2\pi=0.12$
MHz. This nonlinear coupling strength, mode overlap factors calculated
by FEM simulations, and relevant device parameters \citep{Guo2016a}
are used to derive a $\chi^{(2)}$ value of $6.2\pm0.4$ pm/V for
the single-crystalline AlN device platform. This value is quite close
to the 8.6 pm/V recently measured in thin films of MOCVD AlN at 1064
nm \citep{Majkic2017} and is nearly three times higher than previously
reported values for polycrystalline AlN \citep{Graupner1992,Xiong2012b}.

\begin{figure}
\begin{centering}
\includegraphics[width=0.8\columnwidth]{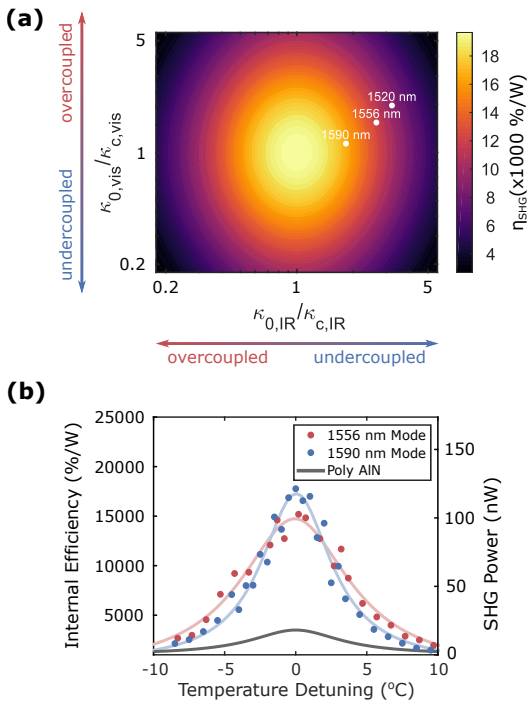}
\par\end{centering}
\caption{\label{fig:Optimal-shg}Optimal coupling and temperature for SHG.
(a) Simulated SHG efficiency from Eq. \ref{eq:eta_full} is parametrically
solved with the cavity coupling rate $\kappa_{c}$ at IR and visible
wavelengths. A peak conversion efficiency of 20,000 \%/W can be achieved
at critical coupling for both modes. Experimental $\kappa_{0}/\kappa_{I}$
are displayed for selected wavelengths. (b) Temperature dependent
SHG efficiency of an optical mode at 1590 nm (blue) and 1556 nm (red).
A Lorentzian fit (solid line) is applied to the experimental data
(circles). A maximum internal SHG conversion efficiency of 17,000
\%/W and 15,000 \%/W are observed for 1590 nm and 1556 nm modes, respectively.
Simulated data from polycrystalline AlN with experimental parameters
from \citep{Guo2016a} is shown for reference (black line).}
\end{figure}

Finally, we look to Eq. \ref{eq:eta_full} to optimize the SHG efficiency.
As noted earlier, the SHG efficiency is optimized when the fundamental
and second-harmonic resonances are both critically coupled. However,
this proves difficult in practice as one (or both) resonances may
deviate from critical coupling when varying the device geometries.
Fig. \ref{fig:Optimal-shg}(a) shows the effect of cavity coupling
on the SHG efficiency (Eq. \ref{eq:eta_full}) using the derived $g/2\pi=0.12$
MHz obtained above and experimentally measured intrinsic loss rates
of $\kappa_{0,IR}/2\pi=240$ MHz and $\kappa_{0,vis}/2\pi=970$ MHz
for infrared and near-visible modes, respectively. The SHG efficiency
is maximized when both modes are critically coupled ($\kappa_{0}/\kappa_{c}=1$)
and decreases as the modes become over- or under-coupled. 

The device used in the previous pump-power experiment in Fig. \ref{fig:Geometrical-optimization}
was nearly critically coupled for the near-visible resonance, but
under-coupled for the IR resonance ($\kappa_{0}/\kappa_{c}>1$), yet
can still achieve a relatively high conversion efficiency of $14,500\pm200$
\%/W. We notice that extinction ratio of the IR resonance increases
when scanning from short to long wavelengths (Fig. \ref{fig:AlN-device-design}(b));
the cavity coupling rate can be increased and a more ideal value for
the SHG efficiency can be obtained by choosing a pump mode at a longer
wavelength. Experimental values of $\kappa_{0}/\kappa_{c}$ are obtained
at various wavelengths and are superimposed on the contour plot in
Fig. \ref{fig:Optimal-shg}(a), which indeed suggest that using a
mode at 1590 nm may increase the experimental SHG efficiency. We systematically
optimize the SHG output of optical modes at 1590 nm and 1556 nm by
varying the global temperature of the chip to align the wavelengths
of infrared and visible resonances, as shown in Fig. \ref{fig:Optimal-shg}(b).
With an input pump power (measured in the input fiber) of 260 \textgreek{m}W,
a maximum SHG output power of 120 nW and 110 nW are collected for
1590 nm and 1556 nm modes, respectively. After considering the transmission
of 22 \% and 10 \% for IR and visible wavelengths, respectively, an
on-chip (internal) SHG efficiency of 17,000 \%/W and 15,000 \%/W can
be calculated for each mode. The observed SHG efficiencies are in
good agreement with the theoretical values calculated in Fig. \ref{fig:Optimal-shg}(a)
for both modes as well as that calculated in Fig. \ref{fig:Geometrical-optimization}(c)
for the 1556 nm mode without fine thermal tuning. We calculate a $\pm$3000
\%/W uncertainty in the calculated conversion efficiency, arising
mainly from the $\pm$10 \% relative error in measuring the absolute
transmission for both infrared and visible wavelengths as well as
amplitude noise from the high-sensitivity visible photodetector. This
internal conversion efficiency is a seven-fold improvement compared
to previous polycrystalline AlN devices \citep{Guo2016a}, manifesting
our result as a new record for SHG among current integrated platforms.
We attribute this record high SHG efficiency to the greatly increased
optical $Q$ factors at both infrared and near-visible wavelengths
compared to previous device platforms as well as a relatively high
material $\chi^{(2)}$ value obtained in this work. We do not believe
a defect-related SHG proces plays a role at this time, as the effective
$\chi^{(2)}$ value for surface charge SHG is typically one to two
orders of magnitude less than that observed in AlN \citep{Levy2011,Puckett2016}.

In conclusion, we demonstrate a fabrication process to realize high
quality AlN-on-sapphire nanophotonic devices for integrated nonlinear
photonic applications. In the low pump power regime, an on-chip SHG
conversion efficiency of 17,000 \%/W (0.17 mW\textsuperscript{-1})
is recorded, which is the new state-of-the-art thus observed among
integrated platforms. At high pump power, pump depletion is observed
at a relatively low pump power of 3.5 mW. The demonstrated high SHG
efficiency paves the way for applications including low loss coherent
frequency conversion \citep{Guo2016b}, on-chip comb self-referencing,
as well as more efficient correlated photon-pair sources \citep{Guo2016c}.

This work is supported by DARPA SCOUT (W31P4Q-15-1-0006) and ACES
programs as part of the Draper-NIST collaboration (HR0011-16-C-0118).
H.X.T. acknowledges support funding from an AFOSR MURI grants (FA9550-15-1-0029),
LPS/ARO grant (W911NF-18-1-0020), NSF EFRI grant (EFMA-1640959) and
David and Lucile Packard Foundation. 

\bibliographystyle{apsrev4-1}
\bibliography{C:/Users/Alex/Documents/BibTex/SHG_Paper}

\end{document}